# Evaluating the impact of government Cyber Security initiatives in the UK

*Adejoke T. Odebade[1], Elhadj Benkhelifa[2]*

[1]*School of Digital Technologies and Art, Staffordshire University, Stoke-on-Trent, UK*

[2]*School of Digital Technologies and Art, Staffordshire University, Stoke-on-Trent, UK*

Corresponding author: *Adejoke T. Odebade (adejoke.odebade@research.staffs.ac.uk)*

**ABSTRACT** Cyber security initiatives provide immense opportunities for governments to educate, train, create awareness, and promote cyber hygiene among businesses and the general public. Creating and promoting these initiatives are necessary steps governments take to ensure the cyber health of a nation. To ensure users are safe and confident, especially online, the UK government has created initiatives designed to meet the needs of various users such as small charity guide for charity organisations, small business guide for small businesses, get safe online for the general public, and cyber essentials for organisations, among many others. However, ensuring that these initiatives deliver on their objectives can be daunting, especially when reaching out to the whole population. It is, therefore, vital for the government to intensify practical ways of reaching out to users to make sure that they are aware of their obligation to cyber security. This study evaluates sixteen of the UK government's cyber security initiatives and discovers four notable reasons why these initiatives are failing. These reasons are insufficient awareness and training, non-evaluation of initiatives to measure impact, insufficient behavioural change, and limited coverage to reach intended targets. The recommendation based on these findings is to promote these initiatives both nationally and at community levels.

## I. INTRODUCTION

In recent years, the UK government has introduced several cyber security initiatives to promote cyber hygiene and address the challenges of cyber threats facing the nation. Promoting these initiatives has been through training and awareness programs, professional certifications and qualifications, competitions, and education. In the latest National Cyber Security Strategy [1], the UK government mentioned the creation of the National Cyber Security Centre (NCSC), launched in October 2016. It was created as a one-stop shop for businesses, the general public, and government, working in partnership with law enforcement, intelligence and security agencies, international partners, and the defence sector. The centre aims to provide incidence response, network security to the public and private sector, information sharing, and minimising the impact of cyber security incidents. The UK government has committed to invest £1.9 billion in cyber security over five years, acknowledging that there are insufficient skills and knowledge in both private and public sectors to meet cyber security needs. As stated in the strategy, the government's vision for 2021 is that "the UK is secure and resilient to cyber threats, prosperous and confident in the digital world" [1]. In the UK, Small and Medium-sized Enterprises (SMEs) make up 99.9% of the business population [2], and a yearly survey carried out by the Department of Culture, Media and Sports (DCMS) showed that micro and small businesses are less likely to seek information and guidance on cybersecurity. It was reported in the 2016 survey [3] that only 2% of businesses sought information from government or public sources. Although this figure increased to 6% in the 2020 survey [4], it is still relatively small.

The impact of cybercrime can be devastating, and cybercriminals are constantly developing hacking methods to obtain information and steal money from regular users, as they are perceived to be the weakest link. Statistics by the Office for National Statistics show that 91% of adults were recent Internet users in 2019, from 90% in 2018. In 2018, the UK was ranked third in Europe for internet use placing UK 10% points above the EU average of 85% [5]. With the increase in cybercrime in the 21st century, it is essential to understand and find ways to hinder the rise in cyber-attacks [6]. In October 2015, the records of TalkTalk, an Internet service provider, were hacked by cybercriminals, which resulted in the loss of sensitive data of over 150,000 customers. The company could have prevented the attack had they taken basic steps to protect customer data [7]. The WannaCry attack on the National Health Service (NHS) in May 2017, which cost the NHS £92m, caused disruptions to the services provided, locking users out of computers and causing more than 19,000 appointments to be cancelled [8]. A document released by the Government on the lessons learnt revealed that eighty of the NHS organisations did not apply basic steps such as patching, and there is also the need to invest more in network security [9]. Considering these events, it is, therefore, crucial that both security professionals and regular users receive training on cyber security [10], [11], [12].

As summarised in Table I below, the UK government has set up initiatives and guidelines to help individuals and organisations protect themselves online.

Table I

UK government Cyber Security (CS) initiatives

| Initiative | Year | Target | Features | Key challenges/concerns |
|---|---|---|---|---|
| Get Safe Online | 2005 | Individuals and businesses | Provides advice on various topics to enable individuals and businesses to protect themselves and their devices from fraud. | Lack of evaluation to measure the initiative's impact on promoting cyber hygiene. Awareness appears heightened during campaign week but is not clear if sustained afterwards. |
| 10 steps guidance | 2012 | Businesses | Ten steps organisations can use to protect themselves in cyberspace | Low adoption of all ten steps, with emphasis more on the technology aspect of guidance |
| Certified Professional Scheme (CCP) | 2012 | Cyber security Professionals | Recognition of competence to cyber security professionals who have continuously applied their skills and expertise to real-life situations. | Cost of application may discourage professionals, thereby limiting those eligible to join the scheme |
| CISP | 2013 | Businesses | A secure environment for exchanging information and increasing situational awareness, thereby minimising the impact of cyber threats | Shared information may include confidential information, and the anonymous sharing of information may create a false sense of security |
| Cyber Essentials | 2014 | Businesses | Provides five security controls to boost cyber security, promotes the image of an organisation and provides a unique selling point | Lack of evaluation to access the breakdown of adoption of controls and the impact of the initiative on organisational culture |
| Cyber Aware | 2014 | Individuals and businesses | Provides advice to individuals and organisations on how to protect themselves online and avoid common cyber problems | Added-value of initiative in mitigating common cyber threats not clear |
| Take-five-to-stop-fraud | 2016 | Individuals and businesses | A campaign that provides advice on how individuals and businesses can protect themselves from financial fraud | High level of fraud despite campaign |
| CyberFirst | 2016 | Young people | Programme to develop young people in cyber security through competitions, courses, and apprenticeships. | Lack of evaluation of the impact on the cyber security profession. |
| Industry 100 | 2016 | Public and private sector | Promotes collaboration between the public and private sectors to address cybersecurity challenges. | Use of feedback to evaluate an initiative may not reliably measure the impact of the initiative |
| Small Business Guide | 2017 | Small businesses | Provides information on how organisations can improve their cyber security | Lack of awareness and adoption. |
| CyBok | 2017 | Academia | Provides a foundation for cyber security education, training, and professional practice. | Regular update of the curriculum is essential due to the dynamic nature of cyber security. |
| Small Charity Guide | 2018 | Charities | Provides information on how charities can improve their cyber security | Lack of awareness and adoption. |
| CyberUK | 2018 | Individuals in charge of cyber security in the public and private sectors. | A national event for the cyber security community to promote relations across various sectors and to discuss the business needs and progressive nature of cyber threats | Possibility of a lack of continuous communication and interrelation after the event. |
| Board Toolkit | 2019 | Board members | Provides resources to enable the board to engage with cyber security | The unclear added value of boards understanding and engagement with cyber security |
| NCSC Certified Training | 2019 | Individuals, training providers | Provides a benchmark for rigorously assessed content and delivery of cyber security training to ensure the quality of the training. | Lack of evaluation may fail to provide the impact of the initiative in addressing the shortage of cyber security professionals |
| Exercise in a Box | 2019 | Businesses | An online tool that helps organisations find out their preparedness and resilience to cyber | Lack of awareness and adoption of the tool |

## II. UK CYBER SECURITY INITIATIVES

### A. CYBER ESSENTIALS

The Cyber Essentials scheme introduced in April 2014 was designed to provide organisations with cost-effective controls to reduce the risks of cyber threats and provide a means for organisations to communicate to their stakeholders the steps they have taken toward cyber security. The 2019 NCSC annual review noted that 14,234 cyber essential certificates were awarded, a 39% increase from the previous year [13]. Although it provides controls to improve cyber security, it does not address advanced or direct attacks against organisations [14]. It was developed in collaboration with the industry and used coaching, documentation, and certification to ensure organisations follow the approach [15]. The scheme provides five self-explanatory security controls, namely, firewalls and gateways, secure configuration, access control, malware protection, and patch management. Although these controls have been designed to help organisations defend themselves against cyber threats, Parkin et al. [16] noted that for Small and Medium Enterprises (SMEs), employees have a joint responsibility in managing security controls whilst bearing in mind the time and skills required to use the controls without interference effectively. They further noted that if the controls are onerous and require a lot of skills, they may interfere with the employees' productivity, and lack of time and skills can lead to a prolonged substandard level of protection.

Beyond the five controls, the Government and industry have worked in partnership to help organisations improve and show their commitment to cyber security by becoming certified. The scheme has two levels of certification: Cyber Essentials, a self-assessment option that helps protect from most common threats, and Cyber Essentials Plus, the same as Cyber Essentials but includes technical verification. Being certified provides an advantage to businesses that want to bid for government contracts that require Cyber Essentials certified status [17]. Tankard [18], however, notes that certification reflects current cyber security practices at the time of assessment. Discussed below are some studies carried out to assess the effectiveness of CE controls and their impact on mitigating threats.

Such et al. [19] investigated the effectiveness of Cyber Essentials controls in mitigating cyber threats by analysing two hundred internet-originating vulnerabilities. The vulnerabilities were randomly selected and analysed across four SME networks with or without the CE controls in place. They employed the use of interviews and surveys, vulnerabilities, and mitigation assessments in carrying out the research. It was observed that 99% of the vulnerabilities were mitigated with the CE tools, but none were mitigated without the CE controls. It was noted that a few vulnerabilities were unmitigated due to outdated or unpatched software or hard-coded flaws in hardware.

Moreover, having CE controls in place does not protect third party networks; for instance, the level of security on a cloud-based service is based on how secure the cloud provider makes it. The analysis of the controls showed that patch management, though least understood by individuals and ranked last in use for SMEs, gave the highest proportion (87.5%) in mitigating vulnerabilities while anti-malware gave the least proportion (10%). Access control, secure configuration, and firewall were rated as the most used controls.

Similarly, Such et al. [20] evaluated the effectiveness of Cyber Essentials in mitigating threats by exploiting vulnerabilities remotely with commodity-level tools by randomly choosing 200 vulnerabilities utilising architectural reviews, configuration reviews, and interviews. Twenty SMEs were surveyed across four sectors, and it was observed that 137 vulnerabilities applied to at least one SME, with code execution, gaining privileges, and denial of service the most common types of vulnerabilities. Their results showed that 69.3% (95) of the vulnerabilities were mitigated using the CE controls, 29.2% (40) were partially mitigated, and 1.5% (2) were not mitigated. The order of effectiveness of the CE controls, according to the study, showed that patch management contributed mostly to mitigating the vulnerabilities, in line with previous research, followed by security configuration, malware protection, firewalls and gateways, and lastly, access control. The study, however, assumed that patches are utilised as soon as they become available, which may not always be the case, and the scope of the study was also limited to commodity-level tools and did not consider other types of threats. It was also observed that some vulnerabilities were partially mitigated due to reliance on patches from third-party vendors and website blacklisting, and the unmitigated vulnerabilities were due to hardware or software flaws.

Parkin et al. [16] used a different model to explore the effect of usability of controls on how individuals can collectively manage organisational security and limit data leakage. The authors modelled the indirect costs of utilising the Cyber Essentials Scheme basic security controls and the Available Responsibility Budget (ARB) to understand how controls can be prioritised for usability and security in small-to-medium enterprises (SMEs). Their results suggested that a combination of anti-malware, two-factor authentication, and correct access privileges give small-sized companies a level of security that minimises the collective burden on staff. It was, however, observed that patch management becomes demanding as an organisation increases in size and complexity.

In a cyber essentials survey carried out by [21] on 251 IT managers, it was revealed that 10% were unsure if their organisation had the certificate, while 19% did not understand the benefit of the certification. Those without the accreditation admitted not having it due to lack of understanding (67%), lack of funds (42%), and not being important (29%). Regarding small businesses (20-49 employees), 33% were either not certified or unsure compared to 6% of larger businesses (750+ employees). While this is a small sample size, it, however, mirrored the Cyber Security Breaches survey over the past five years in fig 2, where micro and small firms are more unaware of the scheme compared to medium and large firms [3],[22],[23],[24],[4].

The findings from the above surveys suggest that the basic controls of the cyber essentials work in mitigating some threats; however, they have been carried out on small sample sizes and SMEs. A more in-depth study is required to assess the effectiveness of the controls on wider threats such as social engineering, physical attackers, insider threats, or other direct attacks [19], various sectors [16], and across large organisations. Though the scheme is effective from the findings, the initiative's impact on behaviour and the extent to which it improves cyber hygiene is unknown; hence, more research is needed to assess the correlation between the initiative and those who have adopted it.

Over the past five years, the Department for Digital, Culture, Media and Sport (DCMS) has been carrying out surveys to help organisations understand the risks they face and how to improve their cyber security status. The number of participants for businesses and charities over the years is shown in Table II.

Table II
Number of participants in cyber breaches survey

|  | 2016 | 2017 | 2018 | 2019 | 2020 |
|---|---|---|---|---|---|
| Micro firms | 278 | 506 | 655 | 757 | 642 |
| Small firms | 174 | 479 | 349 | 321 | 277 |
| Medium firms | 349 | 363 | 263 | 281 | 216 |
| Large firms | 203 | 175 | 252 | 207 | 213 |
| Charities |  |  | 569 | 514 | 337 |

Fig 1 shows the overall awareness of the scheme as reported in the cyber security breaches survey commissioned over the last five years. Charities were surveyed for the first time in 2018, and there has been a yearly increase in the adoption of the scheme across both charities and businesses, which suggests that the scheme is working but rather slowly.

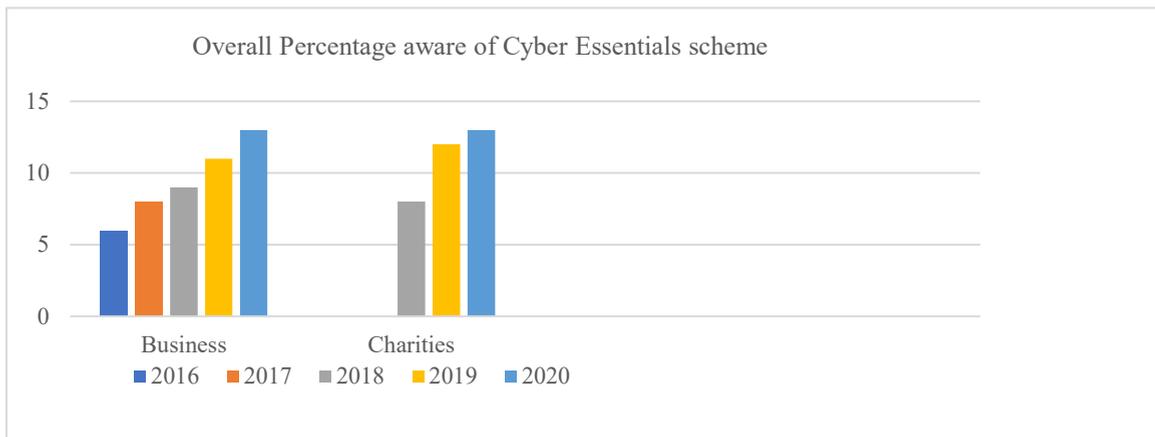

Figure 1: Awareness of Cyber Essentials Scheme

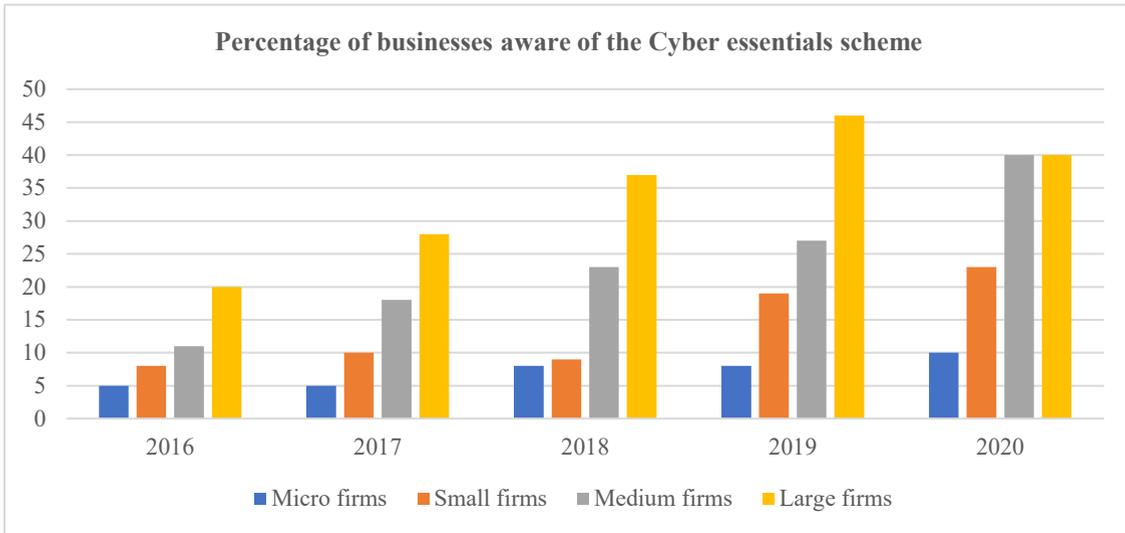

Figure 2: Category of businesses aware of Cyber Essential Scheme

### B. 10 STEPS TO CYBER SECURITY

The guidance was published in 2012 to help organisations protect themselves against cyber-attacks, and it is based on the following 10 steps: risk management, secure configuration, network security, managing user privileges, user education and awareness, incident management, malware protection, monitoring, removable media controls and home and mobile working. It provides resources that allow top-level management to engage with cyber security by producing key questions for senior management to consider in these areas: protection of vital assets, information compromise, and board-level participation [25].

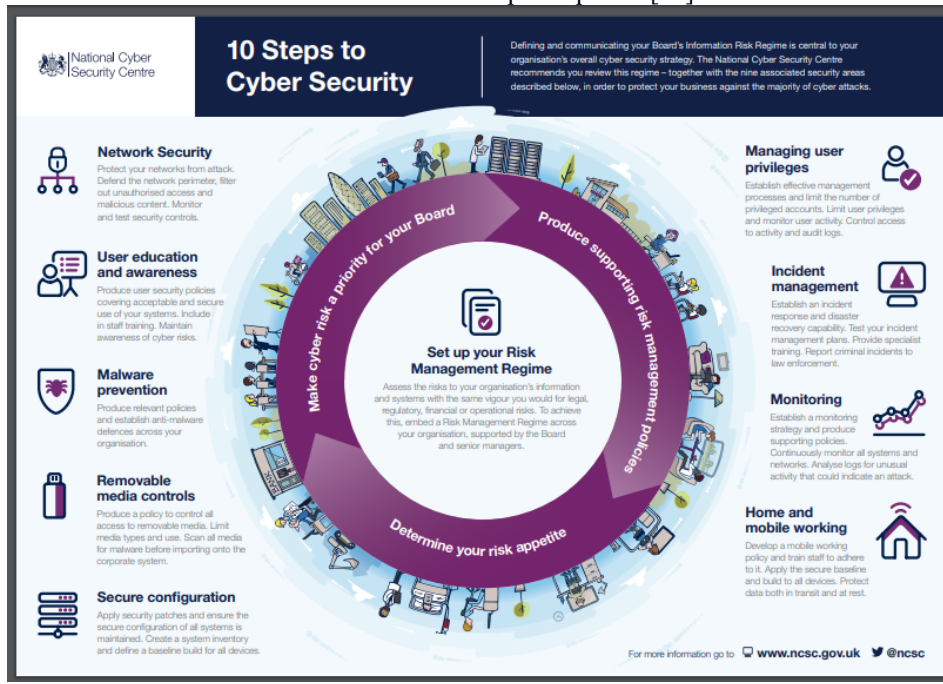

Figure 3: 10 Steps to Cyber Security. Source: National Cyber Security Centre, www.ncsc.gov.uk

Fig 4 and fig 5 below give an overview of the awareness of the guidance over the last five years from the Cyber Security Breaches Survey, showing a greater awareness in businesses and charities than the cyber essentials and a yearly increase in awareness at a slow rate. Hancock [26], the then Minister for Cabinet Office, commented that "Fifty-eight percent of FTSE 350 firms now use our '10 Steps to Cyber Security' guidance, and we've published tailored guidance and free e-learning for SMEs" There is, however, lack of publicly available information that evaluates the effectiveness of the guidance and its impact within organisations.

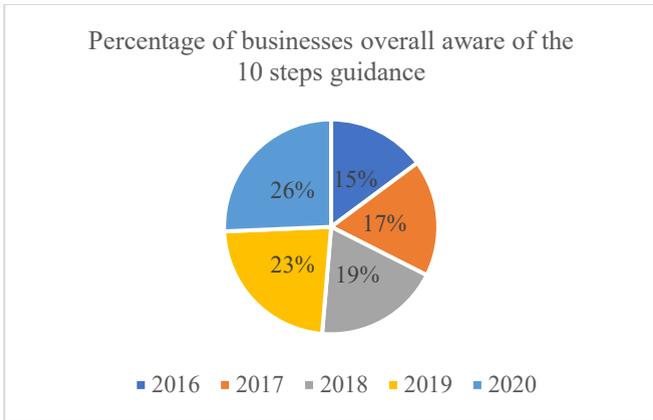

Figure 4: Businesses aware of the 10 steps guidance

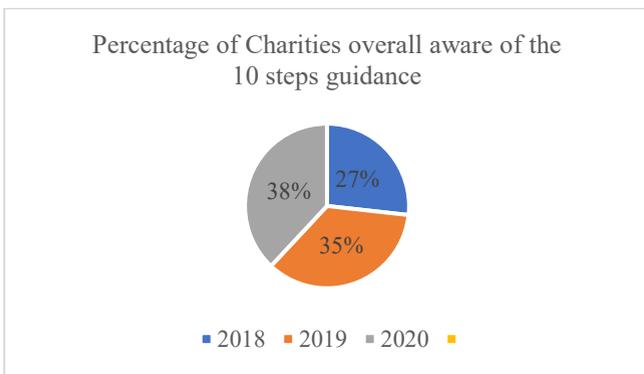

Figure 5: Charities aware of the 10 steps guidance

Fig 6 and 7 show that more than half of businesses and charities surveyed have taken 5 or more of the ten steps, but just a few have progressed in taking all the ten steps, unsurprisingly noticeable in large and medium firms, as shown in fig 8. Looking at the percentage proportion of businesses and charities undertaking action in each of the 10 Steps in tables III and IV below, it is seen that businesses and charities have significantly implemented steps similar to the cyber essentials such as secure configuration and malware protection whereas, actions such as incident management, risk management, and user education and awareness have been less implemented. This indicates that more work is needed to help organisations understand that people and processes are equally important as technology. The 2020 figures in tables III and IV below have been omitted as the questions were worded differently compared to previous years.

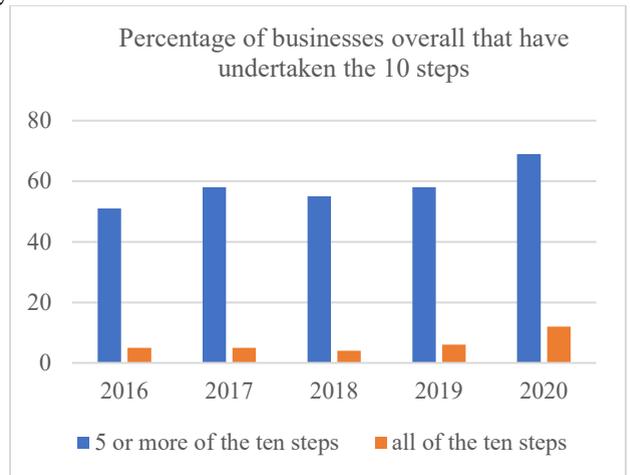

Figure 6: Businesses that have undertaken the 10 steps

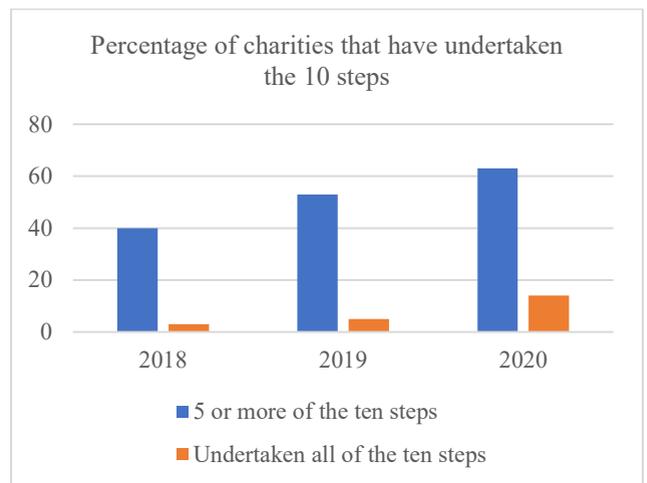

Figure 7: Charities that have undertaken the 10 steps

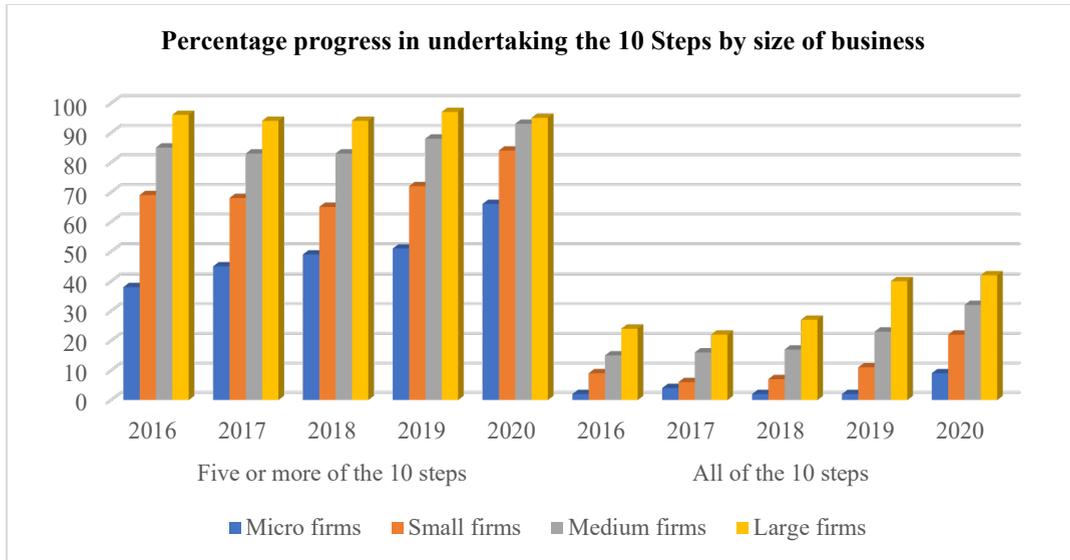

Figure 8: Category of businesses that have undertaken the 10 steps

Table III
Percentage proportion of businesses undertaking action in each of the 10 Steps

|  | 2016 | 2017 | 2018 | 2019 |
|---|---|---|---|---|
| Risk management | 34 | 39 | 35 | 32 |
| Secure configuration | 88 | 92 | 92 | 91 |
| Network security | 86 | 89 | 89 | 89 |
| Managing user privileges | 77 | 79 | 78 | 80 |
| User education and awareness | 28 | 30 | 30 | 37 |
| Incident management | 10 | 11 | 13 | 16 |
| Malware protection | 83 | 90 | 90 | 90 |
| Monitoring | 51 | 56 | 55 | 57 |
| Removable media controls | 21 | 22 | 18 | 22 |
| Home and mobile working | 20 | 23 | 18 | 20 |

Table IV

Percentage proportion of charities undertaking action in each of the 10 Steps

|  | 2018 | 2019 |
|---|---|---|
| Risk management | 27 | 34 |
| Secure configuration | 75 | 80 |
| Network security | 69 | 80 |
| Managing user privileges | 65 | 81 |
| User education and awareness | 21 | 38 |
| Incident management | 8 | 11 |
| Malware protection | 73 | 81 |
| Monitoring | 44 | 73 |
| Removable media controls | 12 | 19 |
| Home and mobile working | 12 | 23 |

### C. CYBER AWARE

Cyber Aware is a national campaign launched in 2017 and offers advice to individuals and organisations on protecting themselves from cybercrime and dealing with common cyber problems. In the light of the recent Covid-19 pandemic, the NCSC launched a new phase of the campaign in April 2020, providing advice on creating separate email passwords, creating strong passwords using three random words, saving passwords in browsers, secure login using two-factor authentication (2FA), software updates and backups [27]. The campaign provides tips to help users develop good online behaviour and change their perspective towards Internet risks. In the 2019 annual review, the NCSC noted its response to producing guidance to users to inform them of an update when WhatsApp, a messaging service, found security flaws in its system. The guidance led to a 54% increase in page views in the first week. Another case study from the review was Black Friday and Cyber Monday, where NCSC published seven tips to guide users when making an online purchase. The podcast was shared about 900 times on social media, liked 2,100 times, and increased the NCSC Twitter followers to over 50,000 [13].

In the last three decades, there has been a huge reliance on digitalisation for data storage, shopping, banking, connecting with others, and much more [28]; however, the impact of cyber awareness in changing people's attitudes online is not certain, for instance, how well do people adhere to creating separate and robust passwords? Do users become aware of the risks of online shopping only when prompted in specific cases, and if yes, is the advice adhered to on an ongoing basis or a one-off?

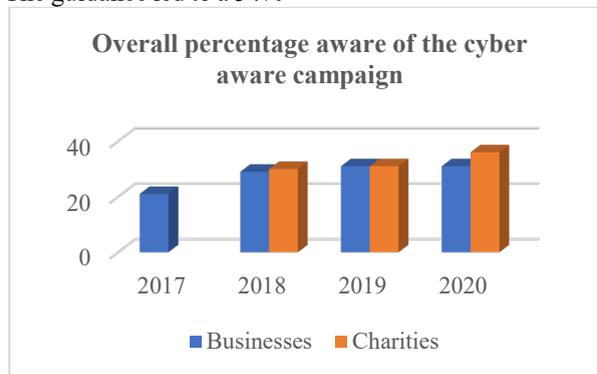

Figure 9: Awareness of Cyber aware campaign

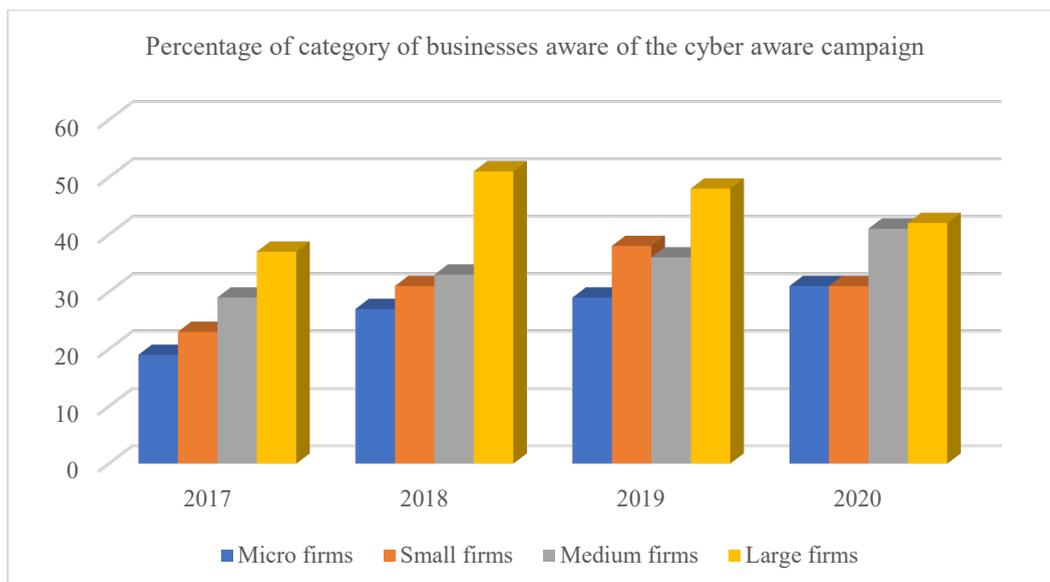

Figure 10: Awareness of Cyber aware campaign by category of business

The awareness of the campaign over the years has increased, as seen in fig 9 above, howbeit slowly, and the figure for businesses has remained unchanged since 2019. Unsurprisingly, large firms are more aware compared to micro and small firms. Although the campaign provides simple and practicable tips that can be used to protect data, it is not sure if these tips reach the general public or if they are aware of the campaign. To the best of the knowledge of this research, there is no publicly available evaluation of the initiative.

### D. CYBER-SECURITY INFORMATION SHARING PARTNERSHIP (CISP)

Cyber security information sharing has been of great interest in the last two decades [29], generating interest in both the public and private sectors, and the synergy between these sectors can lead to profound solutions in addressing the challenges of cyber security [30]. Although information sharing initiatives are commonly recognised, the research into their impact and their role in improving an organisation's position is limited [29]. To encourage information sharing, the UK government launched a Cyber-security Information Sharing Partnership (CiSP) in 2013 as a partnership program between industry and government [31]. Murdoch and Leaver [32] noted that at the end of 2014, the initiative had exceeded its set target, and 777 organisations and 2,223 individuals had joined the program. The NCSC 2019 annual review mentioned that 15,571 members comprise 5,500 organisations from 22 sectors [13]. The initiative was set up to encourage information exchange of cyber threats in real-time, in a secure and confidential environment, providing benefits such as cyber threats warning, learning from the successes and mistakes of others, free network monitoring reports, government-industry engagement, and increased ability to protect the organisation's network. It is a two-step process that requires an organisation to apply for membership, after which staff can then have access to CiSP [33].

Information, as noted by [32], is shared via a secure and restricted access online environment with added features of anonymity and information sharing model, and the information handling model is colour based such that the colour assigned by the author of the information defines how other members of the group further share the information. They further noted that an important part of the program is being able to share information anonymously, which could be due to the nature of the information, legal restrictions, or protecting the image of the business. An initiative such as CiSP provides an avenue for organisations to identify possible vulnerabilities; however, individuals who are technically less proficient may find the information rather technical and complex, making the process complicated, especially when problems are described rather than explained, needing the individual to have proprietary knowledge [19]. While organisations can improve their security and effectively respond to threats by learning from the experience of others, it does come with the issue of trust, as noted by [30], and for organisations to trust the process, it should be transparent and thoughtfully designed to dismiss the fear that information they provide can be misused. They further stated that to gain the trust of users and organisations, those that facilitate the information sharing process should be transparent, skilled, and have a thorough understanding of the objective of the information sharing process. Although information sharing programs provide a platform for organisations to learn from one another and help them be alerted to cyber threats. Little is known as to the number of attacks prevented due to information exchange, how many individuals have been able to better protect their organisation's network or if members of the program have lesser breaches due to early warnings compared to non-members.

### E. GET SAFE ONLINE CAMPAIGN

The Get Safe Online campaign, set up in 2005, is a jointly funded program between the government and private sector that provides an archive of information to individuals and businesses on how to protect themselves online and runs national events such as Get Safe Online week [34][35]. A survey of 1,200 adults by the National Audit Office in 2009 on the Get safe online campaign revealed that only 11% of small businesses and adults were aware of the initiative, and those who explored the site found it useful. A lack of coordinated effort of all Government websites regarding internet security was noted as the link to Get Safe Online; reflected only in six websites out of seventeen. Due to lack of data, it was not clear whether the initiative has led to a change in online behaviour. Although data were collected on the campaign's awareness, it lacked evidence to determine its impact on changing behaviour [34]. Ten years down the line, it appears there is still no evidence to evaluate the impact of the initiative on businesses and the public. A document by The European Union Agency for Cybersecurity (ENISA) with regards to the Get Safe Online Week in 2012 showed an increase of 57% visits to the site compared to the previous week and over 75% increase from referring websites, suggesting that other websites promoted the Get Safe Online website in the context of the event. It was also noted that about 85% of those who visited the site did so for the first time, suggesting the reach of the programme [36], but it is not clear if the increase translated to a change in online behaviour or if it led to continuous visits to the site as was based on one week of the campaign.

Initiatives tailored towards individuals, such as the take five to stop fraud and get safe online, continue to raise awareness about cybercrime and the necessary steps users can take to protect themselves. Although national events such as get safe online week are run to intensify awareness efforts and change online behaviour, there is, however, no evaluation or Key Performance Indicators (KPIs) to suggest whether these initiatives are changing people's behaviour online or whether they are widely known to the general public. It would be useful to be able to determine the following: how many people are aware of the initiative? How many have adopted or used the information provided? How many found it useful? How many unique users and returning users? How much time was spent on the site? How many pages were visited during the duration of the visit? As an initiative that provides information to both the public and businesses, it is imperative to evaluate the effectiveness of the information provided to promote cyber hygiene.

*F. TAKE FIVE*

Take five, a national campaign launched in 2016, is aimed at providing advice to individuals and businesses to be more aware of financial fraud by taking five simple steps: do not divulge PIN or full password, be wary of email and phone call requests, do not feel pressurised to give out information, think through before taking action and stay in control [35]. The initiative has been promoted largely within the financial sector, and according to UK Finance report, the finance sector invested more in security systems last year, thereby preventing £1.8 billion of unauthorised fraud, although £1.2 billion was lost through scams and fraud [37]. The report further revealed that fraud increased, particularly in romance scams, investment fraud, purchase fraud, and money laundering, under the guise of 'Money Mule' adverts. Research by the take five campaign [38] showed that 80% of those surveyed (1,510) would feel ashamed if they were victims of a financial scam, and 68% are more likely to confess if their social media was hacked than admit they fell for a scam (55%).

*G. SMALL BUSINESS GUIDE*

The small business guide was launched in 2017 to enable small businesses to protect themselves from cybercrime by providing advice that businesses can use to improve their cyber security and reduce the risk of becoming a cybercrime victim by following five simple steps: backup data, malware protection, secure devices, password protection and avoiding phishing attacks [39]. The feedback received upon the promotion of the guide led the NCSC to produce a small guide actions leaflet further to make the tips concise and practicable for businesses to implement. Early adopters of the action list surveyed by NCSC found the action list easy to understand and easily integrated into the business [40]. There was no information about the survey conducted; hence the level of awareness of the guide is not known. The guide has been distributed to hundreds of thousands of small and medium-sized enterprises (SMEs) according to the 2019 annual review [13], but it is unclear what percentage of SMEs it translates to. In the 2020 cyber security breaches survey, 16% of micro and small businesses are aware of the guide, indicating low awareness; hence more work is needed to create awareness of the guide.

*H. SMALL CHARITY GUIDE*

Small charity guide provides charity organisations with the information needed to improve their cyber security by providing tips on backups, malware protection, device protection, password protection, and phishing attacks [41]. According to the NCSC annual review, many charities have experienced cyber breaches. However, not all breaches are reported due to fear of reputational damage, and this has led to the design of an education programme tailored towards the charity sector to empower them to protect themselves better [13]. The review further highlights the development of a new program run in local communities by training volunteers to deliver awareness sessions which 96% of participants found helpful in improving their cyber security. An evaluation of such a program is important to assess the effectiveness of the sessions. In the 2020 cyber security breaches survey [4], 16% of charities are aware of the guide. This indicates a low awareness; hence more work is needed to create awareness of the guide. According to the 2020 survey, the target audience of both the small business and charity guides was sometimes unclear as it was sometimes irrelevant for non-technical IT roles and too basic for technical IT or cyber security roles, though useful to educate board members.

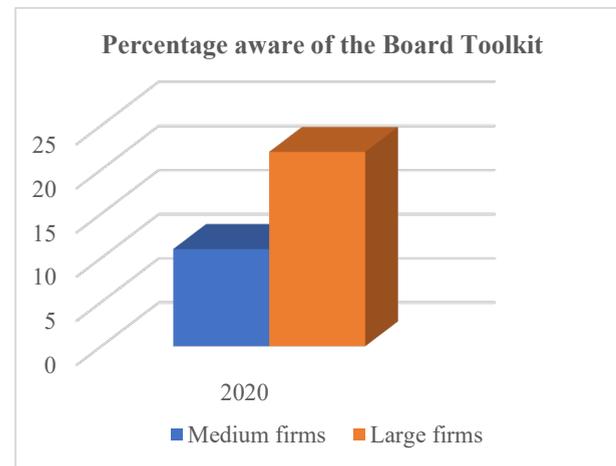

*I. BOARD TOOLKIT*

The board toolkit is designed to help the board engage with cyber security to enable them to make informed decisions, prioritise risks, and manage risks. Board members are central to promoting cyber security in organisations, and while they may not be cybersecurity experts, they must be knowledgeable enough to converse with their experts. The toolkit addresses three questions: What should the board do? What should your organisation do? What does good look like? The kit provides information on an introduction to cybersecurity for Board members, embedding cyber security into structure and objectives, growing cyber security expertise, developing a positive cyber security culture, establishing a baseline and identifying what the board care about most, understanding the cyber security threat, risk management for cyber security, implementing effective cyber security measures, collaborating with suppliers and partners, planning your response to cyber incidents and appendices that summarises legal and regulatory aspect of cyber security [42]. Fig 11 below shows the awareness of the toolkit in the cyber breaches' security survey 2020 by medium and large firms. The survey, however, noted that the findings were broad and did not cover detailed testing, as most interviewees had a glance at the guidance.

Figure 11: Awareness of Board toolkit

Although organisations are becoming more aware of the threats they face in the digital age and are taking steps to protect themselves, the FTSE 350 Cyber Governance Health Check Report revealed that 96% have a cyber security strategy, but only 16% of boards understand the impact of the problems that comes with cyber threats [43]. In response to the report, the then Digital Minister, [44], said: "The UK is home to world leading businesses, but the threat of cyber attacks is never far away. We know that companies are well aware of the risks, but more needs to be done by boards to make sure that they do not fall victim to a cyber attack". Though still in its infancy, an evaluation of the toolkit is important to ascertain if it is helpful to improve boards' understanding of cyber security.

## *J. EXERCISE IN A BOX*

Exercise in a box is a free online tool designed for emergency services, SMEs, and local governments to help them check their resilience to cyber-attacks and evaluate their response. Due to its high demand, the tool is now being used by larger organisations and also outside the UK; hence, it is being reviewed to adapt for the public sector and bigger businesses [45]. George Mudie, the CISO at ASOS, commented in the 2019 annual review that "At ASOS, we decided to incorporate the 'Exercise in a Box' content into our data security incident rehearsal. We found that the desktop exercises and simulation structure really helped to bring the rehearsals to life and encourage discussion and feedback." [13]. Although this tool is relatively new, it however vital to understand if organisations are aware and adopt this initiative.

## *K. CYBERFIRST*

CyberFirst was launched in May 2016 to introduce young people (11-19 years) to cybersecurity and help them explore their passion for technology by providing courses at colleges and universities, bursary and apprenticeship schemes, and girls-only competitions to help tackle gender imbalance in the field, and online courses [46]. The 2019 NCSC annual review revealed that 11,802 girls took part in the CyberFirst girls' competition, and 2,614 students were engaged in the CyberFirst courses leading to a 29% increase in course applications with a 47% increase in the number of female applicants [13]. With the increased level of participation, it is important to be able to able to evaluate, on a longer-term basis, the number of students who eventually take on the cybersecurity profession.

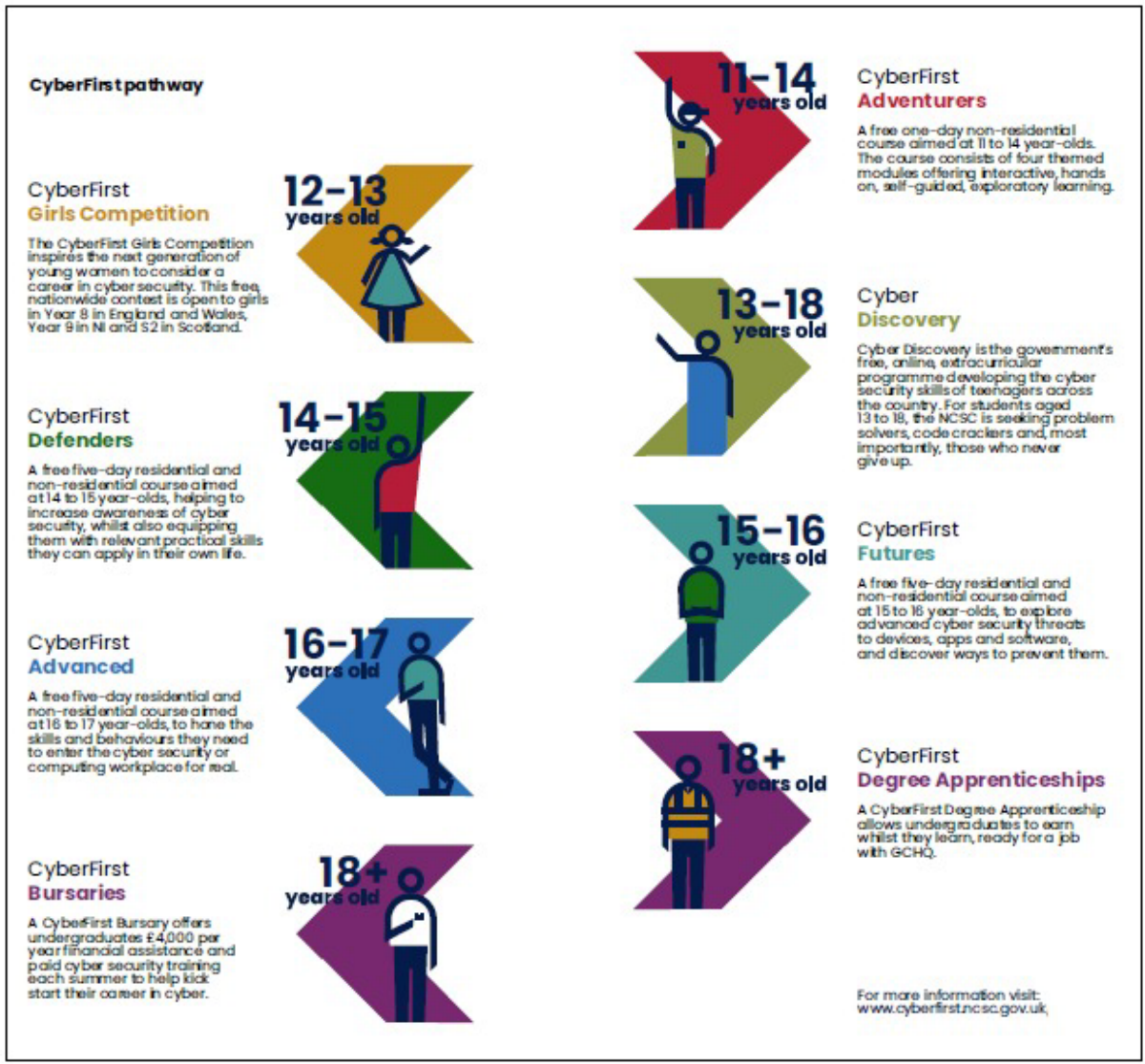

Figure 12 CyberFirst pathway. Source: NCSC Annual Review 2019

*L. CYBER SECURITY BODY OF KNOWLEDGE (CYBOK)*

CyBOK started in 2017 and was launched in January 2020 to systemize the knowledge that underpins cyber security and further aims to develop career pathways and describe contents for courses in higher education and professional training [47]. The CyBOK is expected to bridge the gap in cyber security skills by mapping established knowledge in cyber security [48] in these core areas: Human, Organisational & Regulatory Aspects, Attacks & Defences, Systems Security, Software and Platform Security and Infrastructure Security [49]. The initiative started its first phase in 2017, called Scoping Phase, which involved consultations within and outside the UK, including workshops, surveys, interviews and paper-based exercises, and the second phase, which started in November 2017, saw the creation of 19 Knowledge Areas as shown in fig 13 below which included input, review and feedback by experts and professionals [49]

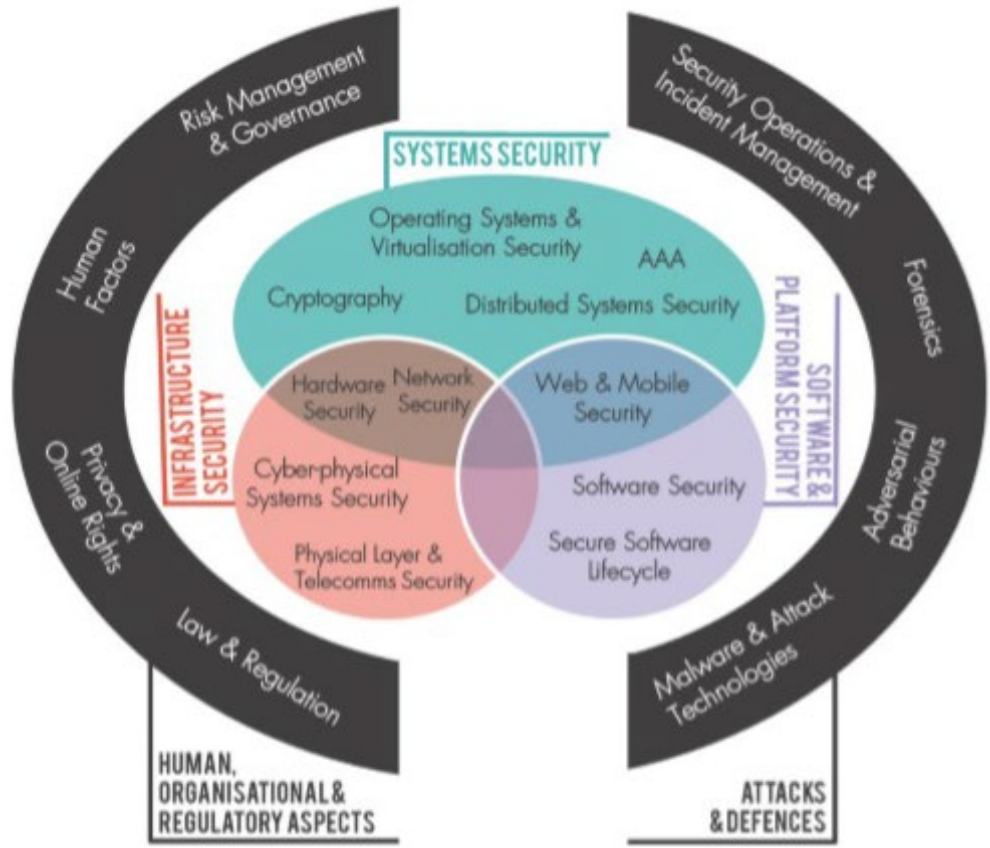

Figure 13: The 19 knowledge Areas (Kas) in the CyBOK Scope. Source: CYBOK

The project is in its third phase to help universities map their cyber security degrees to the NCSC updated programme [48]. A continuous evaluation of the initiative's effectiveness is important to assess whether knowledge gained during the study of the programme is sufficient, transferrable, applicable, and relevant to the industry. A four-week extension for public consultation was published on 3rd July for comments on how to improve the knowledge areas [50].

### M. INDUSTRY 100 (I100)

Industry 100 is an initiative launched by NCSC in 2016 to work with at least 100 industry and government experts. The aim is to create a synergy between the public and private sectors, generating a pool of talents to better understand cybersecurity and ultimately achieve a common goal of keeping the nation safe in cyberspace. The secondees work part-time with NCSC and are paid by their companies to maintain independence [51]. The target was achieved in 2017/2018, and as of the 2019 annual review, 72 more secondees from over 56 organisations have worked at NCSC as part of the scheme [13]. A cyber security consultant [52] shared his experience with NCSC, commenting that "What has impressed me most in my time working on Industry 100 is that representatives from all the different industries (some who are possibly competitors) work together as a team within NCSC in a positive atmosphere towards a common goal. There is a real culture of knowledge sharing and collaboration within the team and a willingness to succeed" [52]. Industries benefit from the initiative by being part of a network of professionals and have the opportunity to contribute to shaping cyber security, brand recognition as part of i100, and a better understanding of government operations [51]. The initiative is evaluated through feedback from secondees; however, it is not certain if there are additional metrics used for evaluation.

### N. CERTIFIED PROFESSIONAL SCHEME (CCP)

The Certified Professional (CCP) scheme is designed to award certification to cyber professionals who have been assessed and verified to have competently applied their knowledge and skills to real-life situations. The scheme provides a community of qualified professionals to work on government networks and a place where employers can look to recruit [53]. The objective of the scheme is to address the shortage of specialists in the profession through consultation with academia, government, and industry, providing benefits to cyber security professionals such as confirmation of professional's competence to deliver business benefits, part of a network of professionals suitable for recruitment as employees or contractors and eligible to work on Critical

National Infrastructure and government networks [54]. Changes were made to the scheme, such as recognising foundational knowledge before applying for specialism, and this is achieved through industry certifications and qualifications that meet certain criteria as this stage is not accessed but rather to ensure professionals have the necessary foundation [55]. Furthermore, another means of recognising foundational knowledge is through the experience-only route for those who do not hold certifications or qualifications within the set criteria but can demonstrate through work experience the knowledge gained using some set areas in the CyBOK [55]. The application cost ranges from £145 to £895 [56], depending on the level of application. The cost may not be affordable for everyone who qualifies to apply, and non-qualification in listed bodies or experience in the identified KA may see fewer applicants to the scheme. To the best of the knowledge of this research, there is no publicly available evaluation of the scheme.

*O. NCSC CERTIFIED TRAINING*

The scheme is designed to provide a standard for cyber security training courses by ensuring the quality of content and delivery; whether for those new to cyber security, that is, awareness level training or professionals looking to enhance their skills, the application level training. There are varieties of certified training courses which are classroom-based, in-house, or online [57]. Training providers go through an assessment process that looks at the provider's quality management, teaching ability and technical knowledge of trainers, with course materials and courses based on the Chartered Institute of Information Security (CIISEC) framework [58]. The benefits of choosing a certified training, as noted by APMG [59], the independent certification body for the scheme, are:
- Delivery and knowledge of trainer have been assessed
- Rigorous assessment of course materials against NCSC's high standard
- Quick identification of high-quality, relevant training by organisations and individuals
- Quality assessment of the course administration process

It is not clear if there is a periodic assessment of training providers after being certified to ensure they adhere to what is expected of the training they provide at the awareness and application levels. An evaluation of the scheme will be essential to assess its effectiveness in addressing the shortage of cyber security professionals.

*P. CYBERUK*

CyberUK is a programme for those in charge of cyber security in government, professionals across industry, the public sector, and Critical Network Infrastructure to foster interaction between the public and private sector through national events where issues such as the evolving nature of cyber threats are discussed to equip individuals and organisations [60] better. The event was attended by 2,767 in 2019 [13], an increase of 11% compared to the previous year, but unfortunately, CyberUK 2020 was cancelled due to the COVID-19 pandemic [61]. There were positive comments from some of the attendees, such as "CYBERUK 2019 was a great opportunity to hear from key influencers in the industry and learn from their experience and best practice. It was a privilege to host the event in Glasgow and a testament to Scotland's commitment to cyber security. To have all these varied stands, speakers and organisations come together is a fantastic opportunity to network and build relationships" [62]. This indicates a welcome approach to enhancing partnerships across different sectors and a platform to gain knowledge. To the best of the knowledge of this research, there are no evaluations to suggest if there is sustained communication and vibe after the event and the impact it has made within the industry.

**III DISCUSSION**

In October 2016, the National Cyber Security Centre (NCSC) was launched with the mission to "help protect our critical services from cyber-attacks, manage major incidents, and improve the underlying security of the UK Internet through technological improvement and advice to citizens and organisations" [63]. The NCSC works with ten Regional Organised Crime Units (ROCUs) that act as points of contact for organisations, businesses, and communities across England and Wales to provide advice on cyber issues [64]. The Government has set up cyber security initiatives. It jointly supports schemes to empower organisations to protect themselves against cyber-attacks. While the tools and resources in these initiatives are not designed to provide a hundred per cent guarantee against cyber-attacks, they help minimise cyber security risks. Hence, the role that people, processes, and technology play in cyber security cannot be overlooked. Technology, the interaction of processes, and users' responses to major security events are all-important cyber security considerations [65].

In determining how secure an organisation is, it is imperative to know if employees are well educated and trained if organisations comply with regulations on protecting and managing data, and if they can measure the security risk of new technologies or services provided [66]. A system can become vulnerable due to a mistake in the design, either deliberately or an accidental omission that questions the system's integrity, confidentiality, and availability [67], and these vulnerabilities can be minimised by paying attention to people, processes, and technology [68].

Lack of understanding, lack of awareness, human behaviour and motive are some of the challenging factors facing cyber security. As pointed out by [69], human mistakes such as leaving devices unattended, sharing usernames and passwords, leaving login details in plain sight, opening links and attachments from unknown sources and negligence are common problems in cyber security. These government initiatives have been designed to circumvent such common problems and to change behaviour, but it is still the case that cyber breaches are linked to these problems suggesting a correlation between human behaviour and cultural change. For cyber security initiatives to be more effective, it is imperative to look at their relationship with behavioural theories such as the Theory of planned behaviour and Theory of reasoned action [70] and Protection Motivation Theory [71], among others.

Changing behaviour requires knowledge and awareness [11] and conscious care behaviour where users consider the effect of their actions, especially on the Internet [72]; therefore, thinking should become the cultural norm in an organisation as a result of inbuilt positive cyber security behaviour [11]. In a bid to promote cyber hygiene, the UK government has set up initiatives and awareness programs to ensure that both home users and businesses feel safe and confident online. Based on publicly available documentation, there are, however, four notable reasons why these initiatives are not meeting their intended objectives, as discussed below:
- A. insufficient awareness and training
- B. non-evaluation of initiatives to measure the impact
- C. insufficient behavioural change
- D. limited coverage to reach the intended target

### A. INSUFFICIENT AWARENESS AND TRAINING

Cybercriminals are constantly developing hacking methods to obtain information and steal money from regular users and organisations; hence, necessary steps should be taken to intensify the awareness of government initiatives to help protect users. While information and advice are available on being secure online, the responsibility lies more on users to source the information. Awareness and training programs are designed not just to provide information to users but to effect changes in people's attitudes and behaviour towards a specific subject, in this case, cyber security. The UK government has developed various cyber security initiatives targeted towards individuals and organisations to improve their cyber security and feel more secure, especially online. While the information and resources available in these initiatives are seemingly simple steps to follow, there is still a lack of awareness.

Given the 1,566 UK organisations (micro-firms through to large firms and charities) surveyed in the 2019 Cyber Security Breaches Survey [24], only 27% of businesses and 29% of charities reported that their staff had cyber security training in the last 12 months. This was expectedly high among the 207 large firms, where 73% responded positively. One of the ten steps to cyber security, user education and awareness, notes the importance of an awareness program in keeping the organisation safe and developing a security culture [24]. However, looking at the number of organisations who took action across the 10 steps in the survey, only 37% responded to having taken action in user awareness and education which is particularly low compared to the technical controls (secure configuration- 91%, network security- 89%, managing user privileges -80%) suggesting that more needs to be done the area of training and awareness.

Increased effort is also needed in the area of 'security by design'. Organisations should be more aware of the need to embed cyber security into every facet of the business, and it should be seen as an ongoing issue rather than a side or one-off. In the latest UK National Cyber Security Strategy, the government stated that "We will build security by default into all new government and critical systems. Law enforcement agencies will collaborate closely with industry and the National Cyber Security Centre to provide dynamic criminal threat intelligence with which industry can better defend itself and promote protective security advice and standards" [1]. Taking a view beyond the UK, the EY's Global Information Security Survey showed that out of almost 1,300 senior leaders surveyed, though a small sample size, only 36% of organisations plan a new business initiative with cyber security from the onset [73].

Although awareness programs help improve cyber security, internet users are faced with challenges such as ignorance of risks of using the Internet, lack of awareness, online-based programs users may find difficult to access due to lack of skills and knowledge, and lack of enforcement [74]. Awareness, as noted by [75], is the bedrock of security culture, and the main aim of awareness in cybersecurity is to effect changes in online behaviour, with security being the focal point [71],[76]. Distinguishing awareness from training, Wilson and Hash (2003) stated that awareness programmes enable users to identify IT security concerns whilst training aims to produce suitable and required IT skills and abilities. Williams [75] pointed out that awareness comprises of education and training; however, Wilson and Hash [77] stated that awareness could underpin a training program. As noted by [10], training should be done regularly with support from management, and factors such as language and culture should be considered when preparing awareness materials. According to Furnell and Vasileiou [78], security is more embraced if staff perceives the message as specific (to them) rather than generic, as the latter is seen as someone else's problem. They further noted that the learning styles of users should also be considered in training and awareness programs.

While there is some information to access the level of awareness on some initiatives geared toward businesses, little is known of those geared towards the general public. To increase the level of awareness, the government should consider raising the awareness of these initiatives at community levels to reach out to more businesses

### B. NON- EVALUATION OF INITIATIVES
The underlying aim of an awareness campaign is to effect a behavioural change [79] which involves a great deal of effort and hence difficult to evaluate due to horizontal and vertical complexity, the unpredictable nature of the intervention, complex and confounding influences, access to appropriate control or comparison groups, lack of knowledge or precision about outcomes and lack of the necessary tools [80]. Aloul [10] conducted a controlled phishing experiment at the American University of Sharjah (UAE) and showed that 954 users out of 11,000 fell victim to a phishing website despite a warning email from the IT department. After the experiment, a website was set up to advise users about phishing attacks. A phishing audit carried out after the awareness sessions revealed that 220 users fell victim to the audit, highlighting the need for a controlled audit in identifying the effectiveness of awareness campaigns. A study conducted by McCrohan et al. [81] to understand problems users have with using strong passwords, for instance, argued that education alone without justification is of minimal value; therefore, training on security practices and user education on e-commerce threats could lead to behavioural change that strengthens online security for both individuals and the organisations they represent [81].

A key aspect in evaluating an awareness program is to determine if people understand the information provided and whether the objective of creating a security culture and behaviour has been achieved as attendance at awareness sessions or the number of leaflets given does not reliably measure the effectiveness of the program [82][83]. Although an increase in site visits, membership, and those who adopt an initiative provides useful information as to the level of awareness of cyber security initiatives, however, it would be interesting and useful to measure the actual behaviour of people towards cyber security and if it significantly improves the security culture in an organisation. A regular evaluation of awareness campaigns is vital so that adjustments can be made where necessary to achieve the desired objective [82]. A possible measurement of behavioural change in cyber security could be through risk reduction [83].

Evaluations have been carried out on some UK initiatives to measure the level of awareness, adoption, or effectiveness, howbeit on a small scale. The evaluations have been geared towards businesses, particularly SMEs, compared to home or regular users. It is uncertain if these initiatives are changing behaviour and organisational culture even among those who have adopted them and if the behavioural change is sustainable. Evaluating these initiatives goes beyond asking people what they think or know; rather, a better understanding is required to know why people behave differently. For instance, what makes an individual ignore an update or click on a suspicious link? Coffman [80] noted that "Just as theory is important for campaign strategy, it is important for campaign evaluation" (Coffman [80]).

### C. INSUFFIEICIENT BEHAVIOURAL CHANGE

In promoting cyber hygiene, various tools and practical advice, such as using a secure password, maintaining regular updates, secure configuration, and installing firewalls and antivirus, among others, have been integrated into cybersecurity practices. Without considering the behavioural changes of users, these pieces of advice and tools may not be as effective as intended. Making cyber security a habit and norm in society may seem a tall order, especially when not enforceable by law compared to driving a car without a seat belt which attracts penalties. Behaviours, attitudes, and knowledge play a significant role in security, especially in the present age of increased cybercrime [84]. Behaviours can be motivated intrinsically, where individuals derive a level of satisfaction and enjoyment for carrying out an activity or extrinsically when individuals engage in an activity because of the reward or the consequences attached to it [71]. Some pertinent questions to consider when developing initiatives or programs strongly linked to human behaviour are: 1) are the initiatives researched-based? 2) are they underpinned by behavioural theories? and 3) on what basis are they created?

With the wealth of resources available to develop good cyber behaviour, the Government office for Science highlights some reasons why people behave insecurely, some of which are: lack of knowledge and skills, no perceived benefit, no perceived risk, attackers use fear and threats and effort required is too high among many others [85].

A survey carried out by Ipsos MORI, commissioned by the National Cyber Security Centre and Department for Digital, Culture, Media and Sport [86], revealed that:
- Only 15% know a great deal about how best to protect themselves from harmful cyber activity
- 46% agree that most information about how to be secure online is confusing
- 34% agree they rely on friends and family for help with cyber security
- 80% say cyber security is a high priority, though this does not mean they take action

The NCSC, in collaboration with Troy Hunt, released the top 100,000 commonly used passwords (NCSC, 2019), with the NCSC director commenting that "We understand that cyber security can feel daunting to many people, but the NCSC has published lots of easily applicable advice to make you much less vulnerable [86]. Hunt also commented that "Making good password choices is the single biggest control consumers have over their personal security posture. We typically have not done a very good job of that either as individuals or as the organisations asking us to register with them" [86]

### D. LIMITED COVERAGE TO REACH INTENDED TARGET

No single factor can mitigate the challenges of cyber threats [87], nor is there a silver bullet in addressing these threats. End users are generally susceptible to cyber-attacks, and in an organisation, employees are considered to be a vital asset;

however, they are perceived to be the weakest link due to the use of the Internet and access to data [87]. Vulnerability in people not only applies to personnel in an organisation but also to home users. While national initiatives and programs in cyber security are great ways for the government to address the challenges home users and organisations face, it appears the information and guidance provided within these initiatives are not reaching the intended target, that is, the general public and organisations.

Looking into the findings of the cyber security breaches survey, specifically where organisations source information on cyber security, as shown in fig 14 below, it is clear that more organisations used external sources than government sources, the main reasons being lack of awareness and misconceptions about government information. This suggests that more effort is required to engage with people and businesses and dispel any wrong notions about government information. Campaigns are more likely to be successful, according to the Government office for Science report [85], if they are complemented with:

- Concurrent community programmes
- Messages being built-in to many different delivery mechanisms
- Role models and champions exhibiting the behaviour
- Policy and law changes
- Readily available products and services to support the target behaviours
- Tailored messages for specific audiences

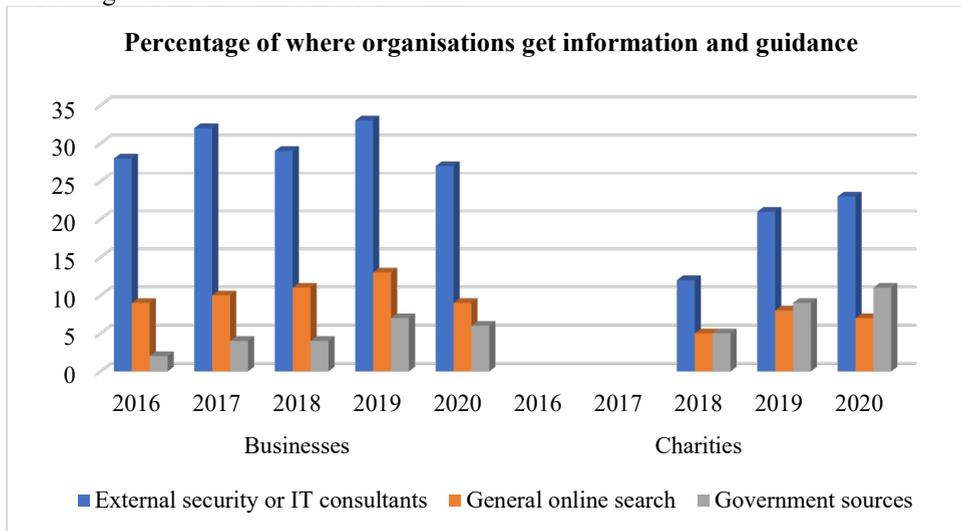

Figure 14: Sources of information for cyber security

The national cyber security initiatives aimed at providing resources needed to stay safe online and changing behaviour can be more effective by engaging local people and businesses within the community, thereby increasing the level of awareness and providing a better means of evaluating the impact on behaviour. Although community engagement is not a silver bullet to the challenges of cyber security, it is, however, a step in the right direction to effectively tackle the challenges of cyber security and foster information sharing among local organisations using national initiatives.

## IV RECOMMENDATION

### A. COMMUNITY LEVEL PROGRAMS

The recommendation for a community-level program is based on the gaps found across the Cyber Security initiatives. While the national initiatives provide effective means to tackle cyber security challenges, it is still the case that the information and guidance provided within these initiatives are not filtering down to the general public and businesses. A contributing factor is lack of awareness which leads to low adoption of these initiatives and will have no impact on changing people's behaviour. A national initiative without enforcement or penalties attached to it or lacks involvement with local people does not resonate with the intended target as sometimes it is taken to be a case of 'I am an unlikely victim' attitude. However, if the awareness of these initiatives is raised with the support of local people and businesses, using real-life scenarios and experiences within the community, the likelihood of acceptance is much higher. There is also a huge disparity in the adoption of these initiatives between small and large firms indicating that more needs to be done to involve local businesses.

One of the ways to disseminate the information and increase awareness is by engaging with people at the community level. One of such is the Community Energy Strategy report by the Department of Energy and Climate [88]. Community energy involves communities getting involved in energy issues in various ways, and this could be in the form of getting better deals on energy tariffs for local people through a local authority's lead in a collective purchasing scheme; local people installing solar systems or wind turbines; community centre running energy advice sessions, among others. The report noted that government alone could not solely tackle challenges. However, involving local people and developing solutions to meet their needs is more effective, thereby creating benefits such as stronger communities, financial benefits for communities, skills, education, and work experience.

Another example is a national training and development programme, called community organisers, funded by the Cabinet Office with the requirement to listen to local people; help people take action on the local issues that are important to them; support people to develop their power to act together for the common good [89]. The findings from the report showed that people living in places where there were community organisers had a higher level of agreement that local people could come together to improve the neighbourhood, and there was also a stronger sense of belonging to their neighbourhoods.

Likewise, there is a need to address the challenges of cyber security and promote national initiatives at the local level and engage local people and businesses within the community. Running cyber security training and awareness programs within the community would create an opportunity to promote information sharing locally and provide a platform for businesses to share experiences that can be of help to other local businesses. People in local businesses can be trained to spot common cyber threats and then take the lead in training and raising awareness of the initiatives. Taking a leaf from national programs run locally as above, running these initiatives will require working in partnership with local authorities as people are more likely to engage because they feel it is targeted towards them, and their needs will be better understood. The likelihood of a sense of community in tackling the issues businesses face in cyber security is high, and information regarding cyber security can easily be dispersed, creating a greater sense of belonging. Running national initiatives locally should also improve the evaluation of the initiatives as the chances of reaching out to businesses and local people is higher.

**APPENDIX**

Table I
Overall Percentage aware of Cyber Essentials scheme

|  | 2016 | 2017 | 2018 | 2019 | 2020 |
|---|---|---|---|---|---|
| **Businesses** | 6 | 8 | 9 | 11 | 13 |
| **Charities** |  |  | 8 | 12 | 13 |

Number of respondents of Businesses (2016-2020): 1008, 1523, 1519, 1566, 1348; Charities (2018-2020): 569, 514, 337

Table II

Percentage of businesses aware of the Cyber Essentials scheme

|  | 2016 | 2017 | 2018 | 2019 | 2020 |
|---|---|---|---|---|---|
| **Micro firms** | 5 | 5 | 8 | 8 | 10 |
| **Small firms** | 8 | 10 | 9 | 19 | 23 |
| **Medium firms** | 11 | 18 | 23 | 27 | 40 |
| **Large firms** | 20 | 28 | 37 | 46 | 40 |

**2016:** 278 micro-firms; 174 small firms; 349 medium firms; 203 large firms; **2017:** 506 micro-firms; 479 small firms; 363 medium firms; 175 large firms; **2018:** 655 micro-firms; 349 small firms; 263 medium firms; 252 large firms; **2019:** 757 micro-firms; 321 small firms; 281 medium firms; 207 large firms; **2020:** 642 micro-firms; 277 small firms; 216 medium firms; 213 large firms

Table III
Percentage of businesses and charities overall aware of the 10 steps guidance

|  | 2016 | 2017 | 2018 | 2019 | 2020 |
|---|---|---|---|---|---|
| **Businesses** | 11 | 13 | 14 | 17 | 19 |
| **Charities** |  |  | 19 | 25 | 27 |

Number of respondents of Businesses (2016-2020): 1008, 1523, 1519, 1566, 1348; Charities (2018-2020): 569, 514, 337

Table IV
Percentage of the category of businesses aware of the 10 steps guidance

|  | 2016 | 2017 | 2018 | 2019 | 2020 |
|---|---|---|---|---|---|
| **Micro firms** | 10 | 11 | 13 | 16 | 18 |
| **Small firms** | 13 | 15 | 15 | 17 | 22 |
| **Medium firms** | 22 | 17 | 24 | 28 | 26 |
| **Large firms**  | 29 | 32 | 34 | 39 | 36 |

Number of respondents of Businesses (2016-2020): 1008, 1523, 1519, 1566, 1348; Charities (2018-2020): 569, 514, 337

Table V
Percentage that has undertaken the ten steps

| BUSINESS OVERALL | 2016 | 2017 | 2018 | 2019 | 2020 |
|---|---|---|---|---|---|
| 5 or more of the ten steps | 51 | 58 | 55 | 58 | 69 |
| Undertaken all of the ten steps | 5 | 5 | 4 | 6 | 12 |
| **Charities** | | | | | |
| 5 or more of the ten steps | | | 40 | 53 | 63 |
| Undertaken all of the ten steps | | | 3 | 5 | 14 |

**2016:** 278 micro-firms; 174 small firms; 349 medium firms; 203 large firms; **2017:** 506 micro-firms; 479 small firms; 363 medium firms; 175 large firms; **2018:** 655 micro-firms; 349 small firms; 263 medium firms; 252 large firms; **2019:** 757 micro-firms; 321 small firms; 281 medium firms; 207 large firms; **2020:** 642 micro firms; 277 small firms; 216 medium firms; 213 large firms

Table VI
Overall percentage aware of the cyber aware campaign

|  | 2017 | 2018 | 2019 | 2020 |
|---|---|---|---|---|
| **Businesses** | 21 | 29 | 31 | 31 |
| **Charities** |  | 30 | 31 | 36 |

Number of respondents of Businesses (2016-2020): 1008, 1523, 1519, 1566, 1348; Charities (2018-2020): 569, 514, 337

Table VII
Percentage of the category of businesses aware of the cyber aware campaign

|  | 2017 | 2018 | 2019 | 2020 |
|---|---|---|---|---|
| Micro firms  | 19 | 27 | 29 | 31 |
| Small firms  | 23 | 31 | 38 | 31 |
| Medium firms | 29 | 33 | 36 | 41 |
| Large firms  | 37 | 51 | 48 | 42 |

**2016:** 278 micro-firms; 174 small firms; 349 medium firms; 203 large firms; **2017:** 506 micro-firms; 479 small firms; 363 medium firms; 175 large firms; **2018:** 655 micro-firms; 349 small firms; 263 medium firms; 252 large firms; **2019:** 757 micro-firms; 321 small firms; 281 medium firms; 207 large firms; **2020:** 642 micro-firms; 277 small firms; 216 medium firms; 213 large firms

Table VIII
Percentage progress in undertaking the 10 Steps by the size of business

|  | 2016 | 2017 | 2018 | 2019 | 2020 | 2016 | 2017 | 2018 | 2019 | 2020 |
|---|---|---|---|---|---|---|---|---|---|---|
| Micro firms  | 38 | 45 | 49 | 51 | 66 | 2  | 4  | 2  | 2  | 9  |
| Small firms  | 69 | 68 | 65 | 72 | 84 | 9  | 6  | 7  | 11 | 22 |
| Medium firms | 85 | 83 | 83 | 88 | 93 | 15 | 16 | 17 | 23 | 32 |
| Large firms  | 96 | 94 | 94 | 97 | 95 | 24 | 22 | 27 | 40 | 42 |

**2016:** 278 micro-firms; 174 small firms; 349 medium firms; 203 large firms; **2017:** 506 micro-firms; 479 small firms; 363 medium firms; 175 large firms; **2018:** 655 micro-firms; 349 small firms; 263 medium firms; 252 large firms; **2019:** 757 micro-firms; 321 small firms; 281 medium firms; 207 large firms; **2020:** 642 micro-firms; 277 small firms; 216 medium firms; 213 large firms

Table IX
percentage of where organisations get information and guidance

|  | Businesses | | | | | Charities | | | | |
|---|---|---|---|---|---|---|---|---|---|---|
|  | **2016** | **2017** | **2018** | **2019** | **2020** | **2016** | **2017** | **2018** | **2019** | **2020** |
| External security or IT consultants | 28 | 32 | 29 | 33 | 27 |  |  | 12 | 21 | 23 |
| General online search | 9 | 10 | 11 | 13 | 9 |  |  | 5 | 8 | 7 |
| Government sources | 2 | 4 | 4 | 7 | 6 |  |  | 5 | 9 | 11 |

**2016:** 278 micro-firms; 174 small firms; 349 medium firms; 203 large firms; **2017:** 506 micro firms; 479 small firms; 363 medium firms; 175 large firms; **2018:** 655 micro firms; 349 small firms; 263 medium firms; 252 large firms; **2019:** 757 micro firms; 321 small firms; 281 medium firms; 207 large firms; **2020:** 642 micro firms; 277 small firms; 216 medium firms; 213 large firms